\documentclass[twocolappendix,  twocolumn]{openjournal}

 \usepackage{amssymb}
 \usepackage{natbib}
\usepackage{graphicx}
 \usepackage{lineno}
 
\usepackage{xcolor}
 
\usepackage[utf8]{inputenc}
\usepackage[english]{babel}
\usepackage{hyperref}
\hypersetup{
    colorlinks=true,
    linkcolor=blue,
    filecolor=blue,      
    urlcolor=red,
    citecolor=blue,
}
\usepackage{color,colortbl}

 \DeclareGraphicsExtensions{.bmp,.png,.jpg,.pdf}
\usepackage{verbatim}
\usepackage[normalem]{ulem}
\usepackage{orcidlink}
\usepackage{soul}

\begin{document}

   \title{ small scale turbulence alongside with large scale turbulence  in  a z=1.87 star Forming Galaxy with outflowing wind, revealed by multi-point structure functions }
\shorttitle{small scale turbulence}
\author{Itzhak Goldman $^{1,2}$}
\affiliation{$^{1}$Physics Department, Afeka College, Tel Aviv 6998812, Israel}
\affiliation{$^{2}$ Astrophysics Department, Tel Aviv University, Tel Aviv 6997801,  Israel}

\begin{abstract}
  Recently,  \citet{Goldman2024} used observations by \citet{Vasan+2025}  and  obtained evidence for a large scale compressible, Burgers  turbulence in the ISM of  CSWA13, a  gravitationally lensed,  star-forming galaxy at $z = 1.87$, with an outflowing wind,
 
  \citet{Goldman2024}  analyzed the residual line of sight velocity field of the nebular gas, derived from by the   C$_{|||]}$ emission line, and the residual, line of sight velocity field of the   outflowing wind  obtained by observations of the    Si$_{||}^*$ florescent emission  line. The two velocity fields are functions of the position  along the major galactic axis, 

The turbulence timescale on the largest spatial scale has been found to be
$\sim$ 500 Myr . This together with the large spatial scale of $\sim$6.43 kpc  suggest a large scale generating mechanism  (such as tidal  interaction or merger) that lasted
for  a time $\sim$500  Myr . On the other hand, the outflowing wind  is much younger and is probably the result of the intense star formation. Therefore, could it be that the star formation  drives also turbulence on small scales?

In the present paper we utilize multi-point second order structure functions \citep{Cho2019, Seta+2023, Seta+Federrath2024, Lee+2025} to find whether there exists also a small scale  turbulence in this galaxy, and if so, try to identify its drivers.

We obtained evidence for small scale turbulence whose largest spatial scale is $l_s \sim$240  pc  for the nebular gas velocity field and  $l_s \sim$290  pc  for the outflowing wind velocity field.  These values suggest that   stellar sub clumps or large  star clusters with an high concentration of young massive stars  could be   responsible for both the outflow and for the small scale turbulence. 

In the case of the wind velocity there is an additional,  small scale turbulence, with $l_s \sim$1.5 kpc. This is about the size of the large clumps of that galaxy, suggesting these large clumps as the drivers of the turbulence on this scale. The timescale of the small scale turbulence is $sim$127 Myr for the nebular wind velocity, and $sim$1155 Myr and$sim$162 Myr for the wind velocity. These values are  in accord with a late formation, compared to the large scale turbulence. 
\end{abstract}
 \maketitle

\section{introduction}

High redshift galaxies are characterized by high star formation rates as well as  outflowing winds, generated by the young stars or by AGNs, e.g. \citep{Bournaud+2009, Hoffmann+2022, Rizzo+2021, Sanders+2023, Shah+2022}.  The high rate of star formation is attributed to the assembly process of the galaxy, The gas supply can be in the form of  inflow from the circum galactic medium (CGM) and also by more violent events  such as   mergers and tidal interactions.   

 Observations of high redshift galaxies  display velocity dispersions that are usually interpreted as manifestation of turbulence e.g. \citet{ Burkert+2010}. It has been argued that accretion onto disk galaxies can generate large scale turbulence, in particular at the disk outskirts, e.g. \citep{ Forbes+2023, Goldman+Fleck2023}. Turbulence can be generated also by mergers and tidal interactions. To establish the existence of turbulence and moreover, to understand its nature, a power spectrum or structure function  of the velocity field are  needed. This in turn, demands observations with high enough spatial resolution which, for galaxies at high redshifts, are challenging.
 
Gravitational lensing can help in this regard.     A  recent paper   \citep{Vasan+2025} presented a study of a wind outflow in CSWA13,  which is   a gravitationally lensed star-forming galaxy at $z = 1.87$. The gravitational lensing allowed for a substantially  improved spatial and kinematic resolution. The   authors
     obtained, among other results,  two line of sight  velocity fields along the major axis of the galaxy:  the nebular gas velocity traced by the  C$_{|||]}$ emission line, that represents also the velocity of the young stars embedded in the nebular gas (as was validated by \citet{Rhoades+2025}), and the outflowing wind velocity traced by the Si$_{||}^*$ florescent emission  line. The semi forbidden    C$_{|||]}$ emission line is emitted in the Worm ionized Medium (WIM) of the interstellar Medium (ISM). The    Si$_{||}^*$ florescent emission  line is emitted by the WIM as well as by Photo Dissociation Regions (PDRs).
     
  In  \citet{Goldman2024}, we used the residual velocity fields (after subtracting a large scale gradient and the remaining mean) and computed power spectra and 2 point second order structure functions. The latter indicated the existence of a large scale ($\sim$ 6.43 kpc) compressible, Burgers turbulence. The timescale of the turbulence on the largest spatial scale turned out to be $\sim$ 455  Myr  for the nebular gas velocity field, and  $\sim 523 $ Myr for the outflowing wind velocity field. The large spatial scale of the turbulence as well as the large timescales suggest that  the generating mechanism of the turbulence is a large scale one: a merger or tidal interaction with another galaxy. 

This galaxy is forming stars, and the lensed images show the existence of stellar clumps. It has been argued that high z galaxies are clumpy with the clumps containing molecular  gas where stars formation takes place. The outflowing wind   probably  is  a feedback from  the intense star formation.

In the present paper we investigate wether the intense star formation   also drives  a small scale turbulence.
We employ multi-point second order structure functions  \citep{Cho2019, Seta+2023, Seta+Federrath2024,  Lee+2025}. A break in the slope of a multi-point structure function at a small spatial scale, $l_s$,  indicates the existence of a turbulence for which this small scale is the largest spatial scale.

 \citet{Cho2019} explains that when the break is followed by a plateau, the case for the existence of small scale turbulence is stronger and the value of the structure function at the plateau equals twice the square of the standard deviation of the small scale turbulent quantity; in the present case the turbulent velocity. velocity.
 
 \begin{equation}
 SF(l_s)= 2 \sigma_s^2
 \end{equation}
where  $\sigma_s$ is the  standard deviation of the small scale turbulent velocity. This follows from the fact that the auto correlation at a lag equal to $l_s$ is zero.

  \citet{ Lee+2025} applied multi-point structure functions to analyze the HI intensity maps of the Small Magellanic Cloud (SMC) and found small scale turbulence with  $l_s \sim 50$pc suggesting the HI shells as drivers of the turbulence. In addition, there exists also a large scale  turbulence with a largest scale  of $\sim 3$kpc  that was probably generated by a tidal interaction with the Large Magellanic Cloud (LMC) about 200 Myr ago \citep{Goldman2000}.
  
 The paper is organized as follows:In section 2, the multi-point  structure functions used, are introduced. in Section 3. the data from \citet{Vasan+2025}  used in \citet {Goldman2024} and in this paper, is presented. In section 4, the multi-point structure functions of the residual  nebular velocity field  are computed.  In section 5, the multi-point structure functions of the residual  velocity field of the outflowing  wind are computed. Discussion is presented in section 6. 
\section{Multi-point second order structure functions} 

The familiar 2 point structure function of a quantity $f(x)$ is defined as

\begin{equation}
SF_{2pt}(x)=<| f(x+x')- f(x')|^2>.																											
	\end{equation} 
The angular brackets represent averaging over $x'$. 
  
The definitions of the 3 point, 4 point, 5 point and 6 point   structure functions are  \citep{Cho2019}:
\begin{eqnarray}
  SF{3 pt}(x)=\frac{1}{3}<|f(x'-x) -2 f(x')+f(x'+x)|^2>.\\
   SF_{4 pt}(x)=\frac{1}{10}<\vert f(x'-x) -3 f(x')\nonumber  \\ 
+3f(x'+x)  - f(x'+ 2 x )\vert^2>.\\																		
 SF_{5 pt}(x)= \frac{1}{35}<\vert f(x' - 2 x) - 4 f(x'-x) +6 f(x') \nonumber \\
 -4 f(x' +x) +f(x'+2 x)\vert^2 >.\\
 SF_{6 pt}(x)= \frac{1}{126}<\vert f(x' - 2 x) -5 f(x'-x)+ 10 f(x') \nonumber \\
 -10 f(x'+x)+5 f(x'+2 x) -f(x'+3 x)\vert^2>,
 \end{eqnarray}

Here too, the angular brackets denote averaging over $x'$.

\section{Data}

\citet{Goldman2024} digitized the velocity curves  of Figure 8 of \citet{Vasan+2025}                                                                                                                                                                                                                                                                               and obtained the line of sight nebular and wind velocity as functions of position along the galactic major axis. For the sake of clarity, we summarize below the steps that led to the residual nebular and outflowing wind velocities in \citet{Goldman2024}. We note   that \citet{Vasan+2025}                                                                                                                                                                                                                                                                   
did not provide uncertainties for the two velocity fields, 
 
\subsection{The residual nebular velocity along the  galaxy major axis}
  
\begin{figure}[ht!] 
  \centering
   \includegraphics[scale=0.3] {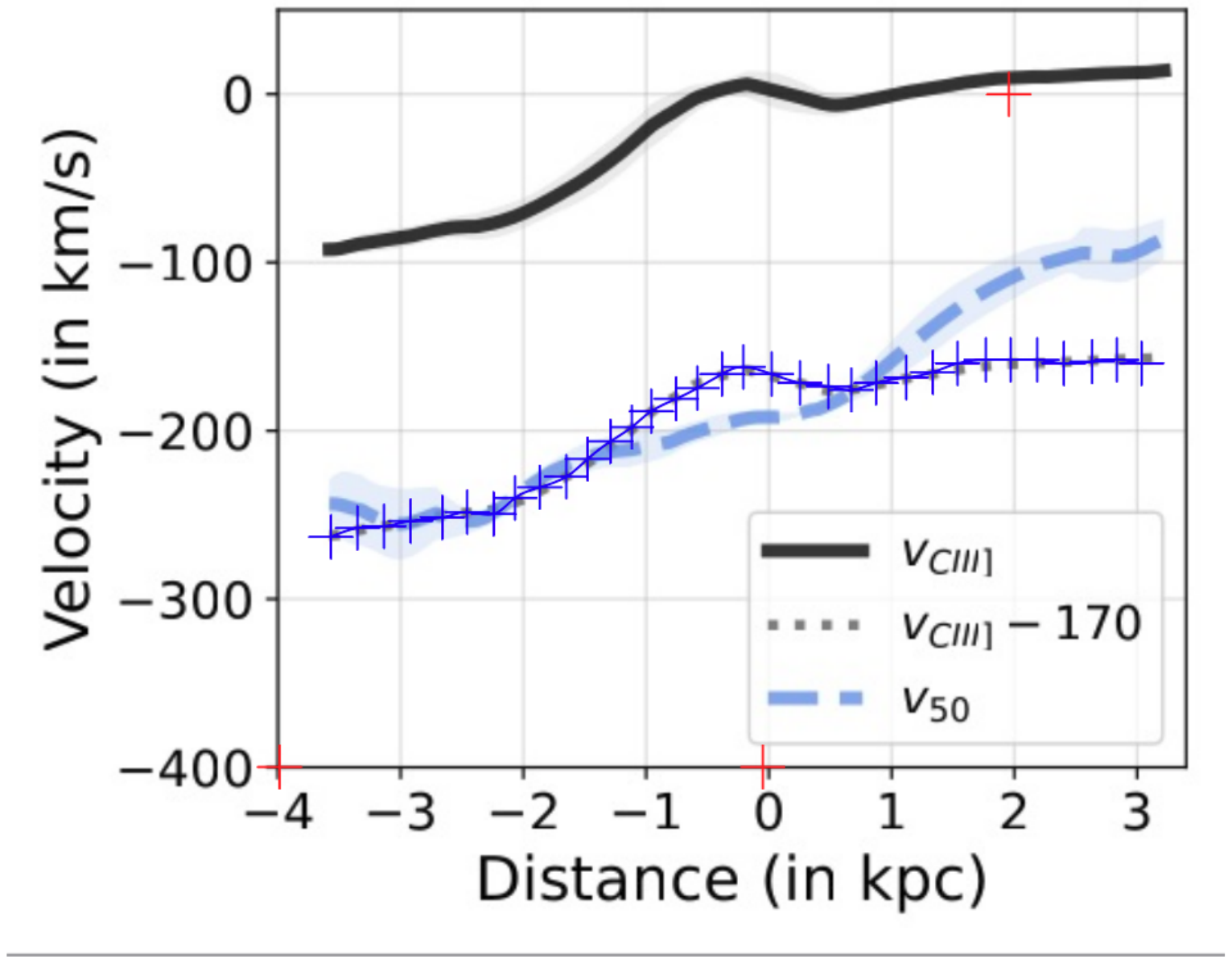} 
  \caption {The marked nebular velocity, offset by 170 km/s, in units of $km/s$ as function of position along the major axis, in units of $kpc$.}
  \label{v_nebdig}
       \end{figure}

The Engauge Digitizer Ver.12.1, has been used to mark and read   the nebular velocity of Figure 8  of \citet{Vasan+2025}.
The marked observed nebular velocity,   offset by 170 km/s,   is shown in figure \ref{v_nebdig}.

\begin{figure}[ht!] 
  \centering
   \includegraphics[scale=0.4] {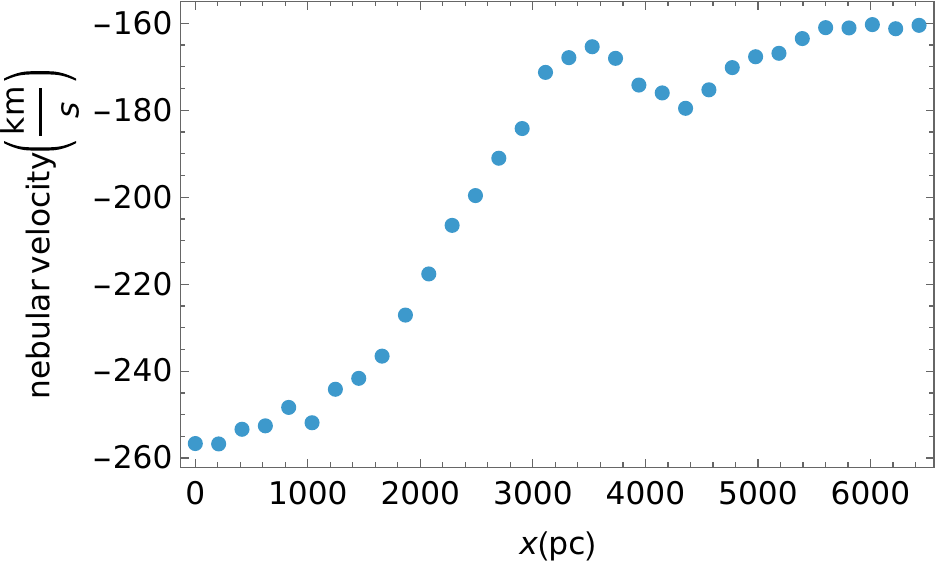 }
  \caption {The  nebular velocity in units of km/s as function of position along the major axis, in units of   pc.}
  \label{v_neborig}
       \end{figure}

    The Engauge Digitizer yielded the values of the velocity and position at the marked points as seen in figure \ref{v_neborig}. The spacing between two adjacent marks is 207.5 pc.
    
\begin{figure}[ht!] 
  \centering
   \includegraphics[scale=0.4] {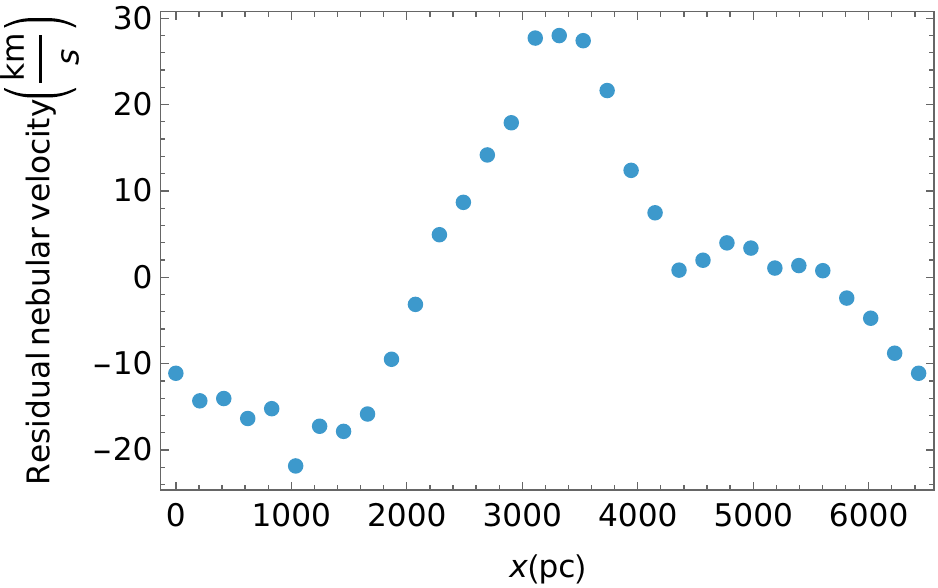 }
  \caption {The residual nebular velocity in units of km/s as function of position along the major axis, in units of  pc.}
  \label{vneb}
       \end{figure}

 The nebular velocity posses a  large scale gradient of  96.2 km \ s$^{-1}$/ ( 6.43 kpc).   After subtracting the gradient, and the remaining mean value, the residual velocity is obtained and is displayed in figure \ref{vneb}. We emphasize that the above figures displays the velocity at the discrete positions, that are spaced from each other by $207.5$ pc.

\subsection{The residual wind velocity along the  galaxy major axis}
  
\begin{figure}[ht!] 
  \centering
   \includegraphics[scale=0.3] {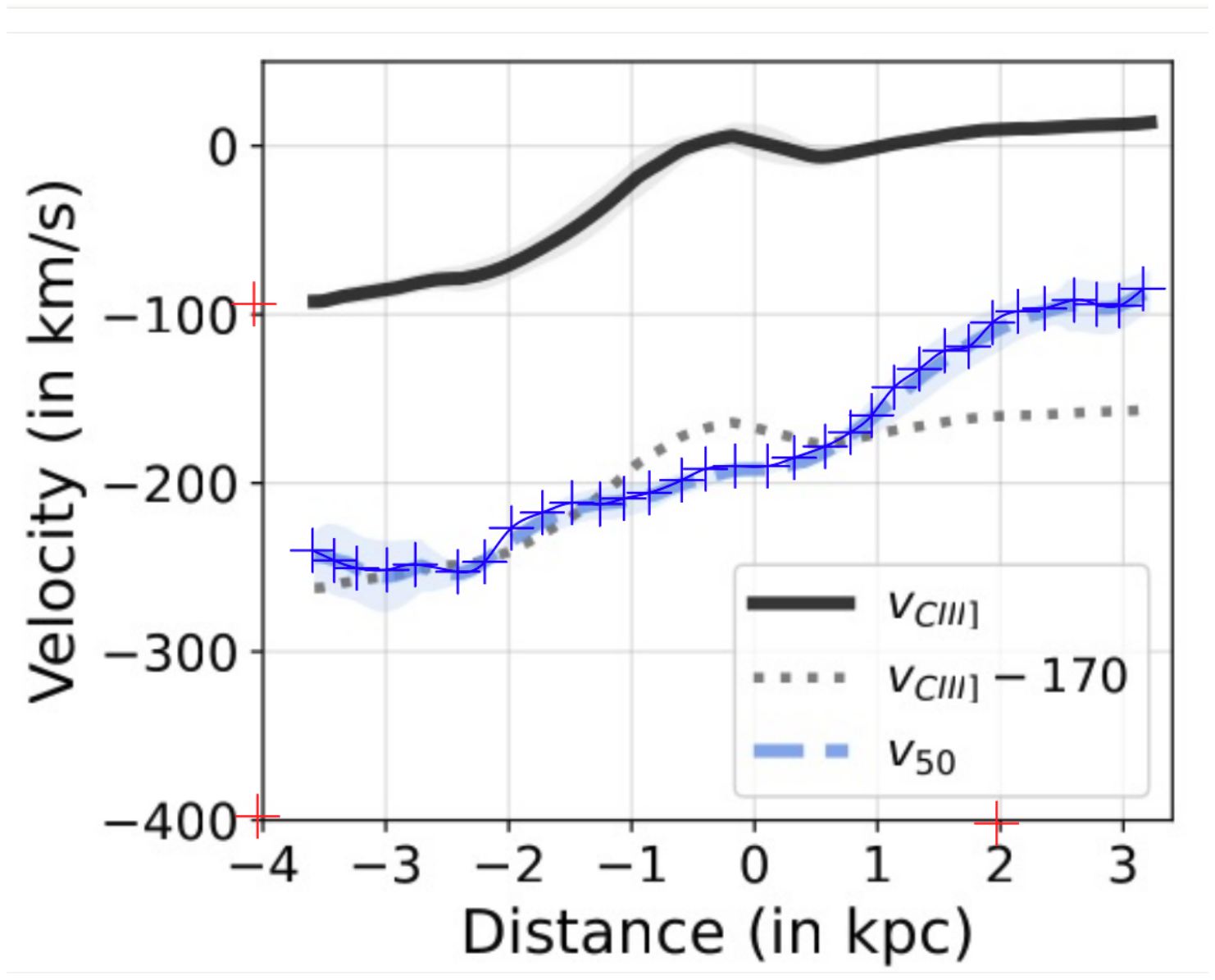} 
  \caption {The marked  wind velocity in units of $km/s$ as function of position along the major axis, in units of $kpc$.}
  \label{v50dig}
       \end{figure}

 We followed similar   steps in treating the wind velocity. The marked digitized wind velocity of Fig. 8 of \citet{Vasan+2025},  is displayed in figure \ref{v50dig}. The spacing between two adjacent marks is 207, 5 pc.
  
   \begin{figure}[ht!] 
  \centering
   \includegraphics[scale=0.4] {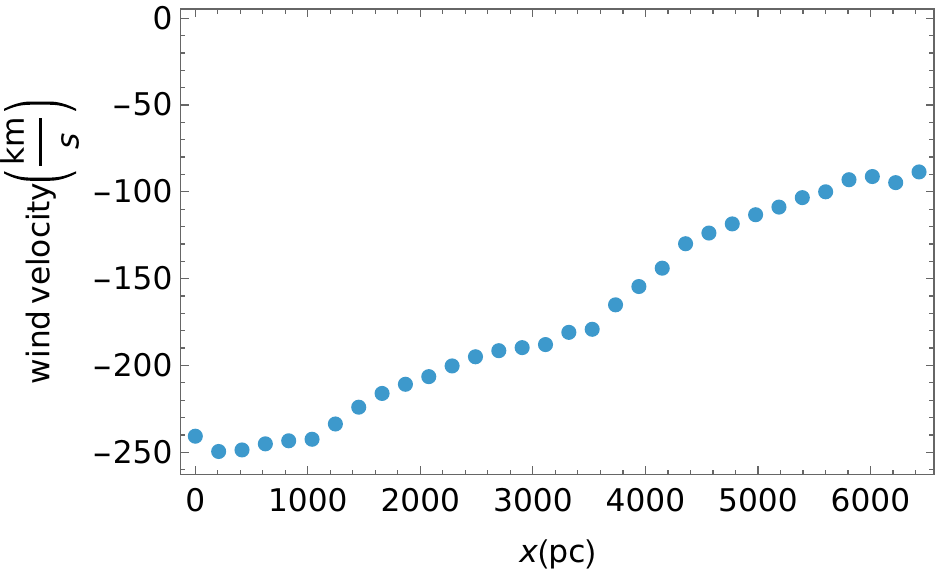}
  \caption {The  wind velocity in units of km/s as function of position along the major axis, in units of   pc.}
  \label{v_50orig}
       \end{figure} 
    The Engauge Digitizer yielded the values of the velocity  and position of the marked points as seen in figure \ref{v_50orig}.
     
\begin{figure}[ht!]
  \centering
   \includegraphics[scale=0.4] {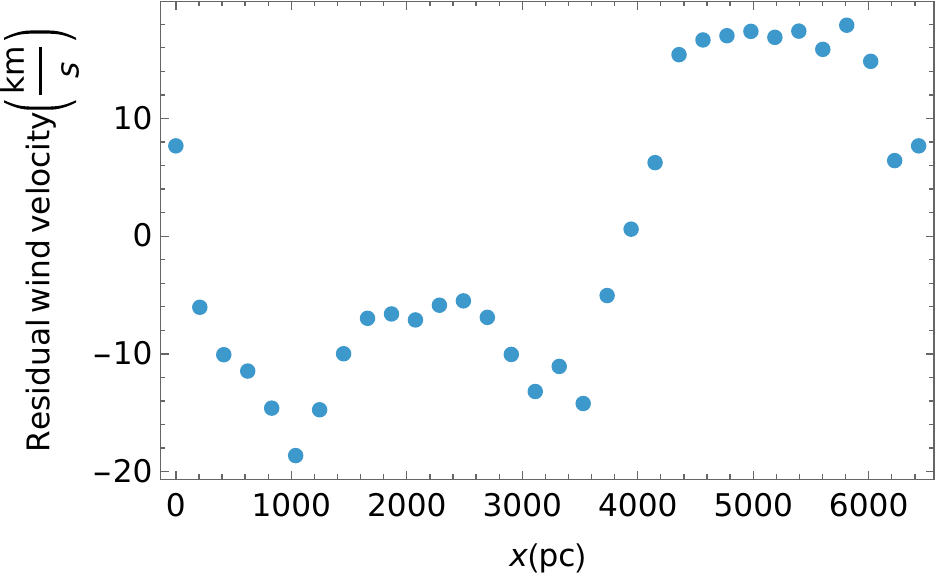}
      \caption{The residual wind velocity in units of km/s as function of position along the major axis, in units of  pc.} 
             \label{v_50}
                    \end{figure}

    The wind velocity posses a  large scale gradient of  152.2 km s$^{-1}$/( 6.43 kpc).   After subtracting the gradient, and the remaining mean value, the residual velocity is  displayed in figure \ref{v_50}.
    
\section{Multi-point structure functions of the residual nebular gas velocity field.}

\begin{figure}[h!] 
  \centering
   \includegraphics[scale=0.4] {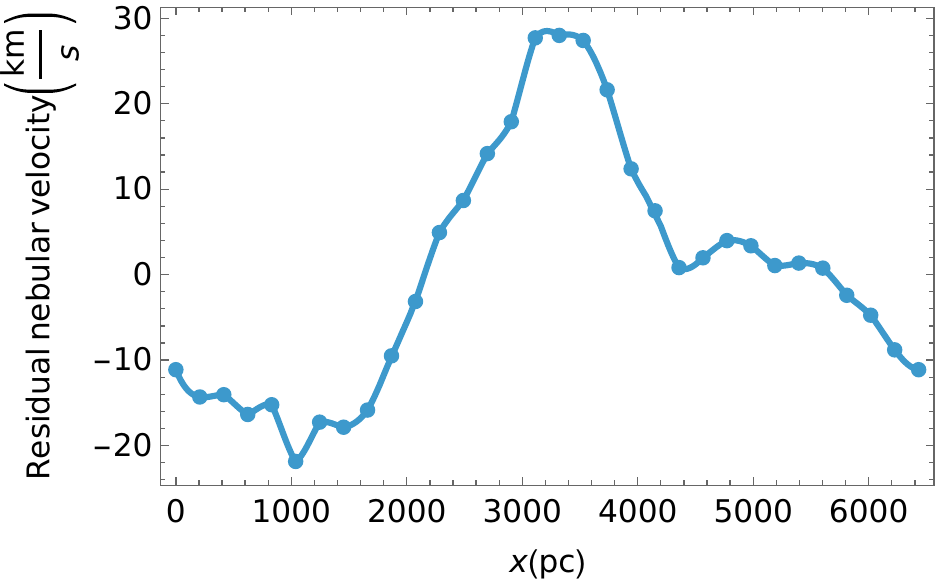 }
  \caption {The residual nebular velocity field, and an  interpolation  fit in units of km/s  as function of position along the major axis, in units of  pc.}
  \label{v_neb_fit}
       \end{figure}

We used Wolfram Mathematica 14.3 to obtain an   interpolation fit to the residual discrete velocity field of the nebular gas,  figure \ref{vneb}.  The discrete  and the   interpolation  fit velocity fields   are shown in figure \ref{v_neb_fit}. 
 
  The averaging  that appears in the definitions of the multi-point structure functions is performed by integrating over the entire 
 x-span of L= 6.43 kpc and division by L.

\begin{figure}[h!] 
  \centering
   \includegraphics[scale=0.55] {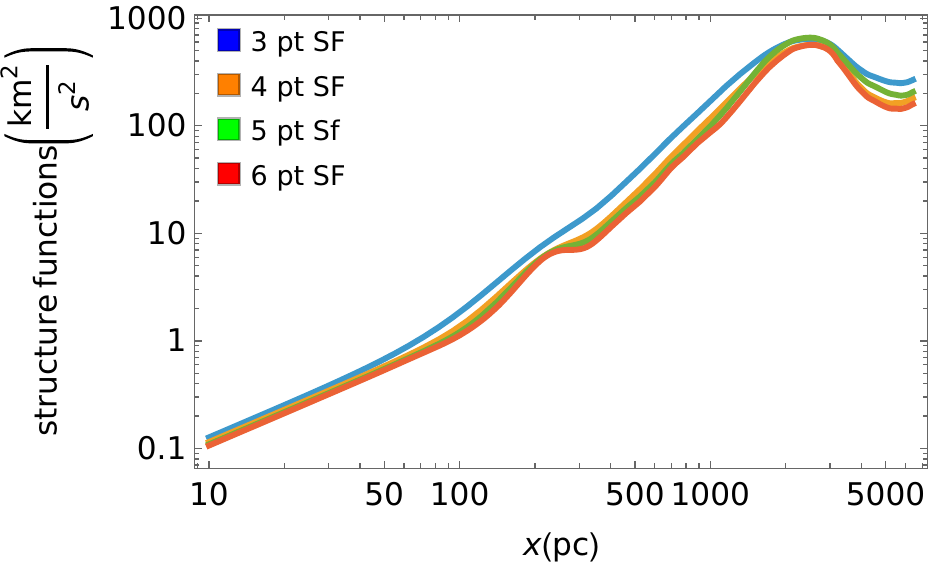 }
  \caption {Multi-point structure functions of the residual nebular gas velocity field, in units of  
  (km/s)$^2$ as function of spatial lags along the major axis, in units of    pc.}
  \label{SFs_v_neb} 
  \end{figure}
 
  The  computed multi point structure functions of this velocity field  are shown in  figure \ref{SFs_v_neb}.

 Inspection of figure \ref{SFs_v_neb} reveals that the 3 point structure function shows only a hint for the small scale turbulence. The 4, 5 and 6 point structure functions all show a common plateau.
  \citet{Cho2019} noted that when the multi-point structure function involves more points the plateau is clearer.  
 
The  plateau in figure \ref{SFs_v_neb} begins  at $l_s\sim$ 240 pc and has a width of $\sim$ 120~pc.

 The corresponding value of  the structure functions is 
 \begin{equation}
 SF(l_s)= 2\sigma_s^2= 6.9\  km^2/s^2,
 \end{equation}
 
 implying    $ \sigma_s$=1.86 km/s. 

 This is the value of the turbulent velocity on the scale $l_s$ which is the largest scale of the small scale turbulence. The associated  timescale is $l_s/ \sigma_s\sim$ 127 Myr.

 \section{Multi-point structure functions of the residual outflowing wind velocity field.}
  
\begin{figure}[h!] 
  \centering
   \includegraphics[scale=0.4] {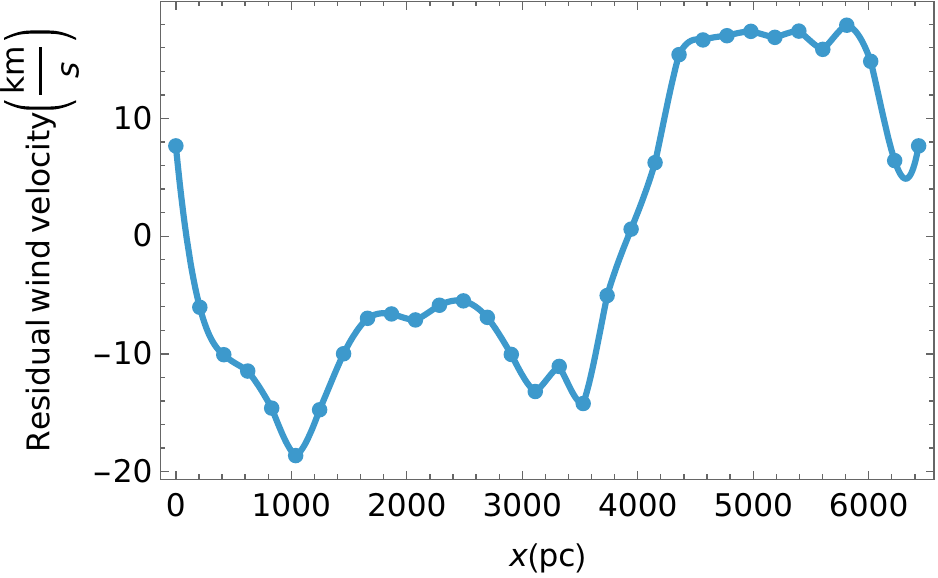 }
  \caption {The residual outflowing wind velocity field, and an interpolation fit in units of  km/s  as function of position along the major axis, in units of pc.}
  \label{v50_fit}
       \end{figure}

 Here too, Wolfram Mathematica 14.3 was used to obtain an interpolation fit  to the residual discrete velocity field of the outflowing wind (figure \ref{v_50}).  The  discrete and the  interpolation fit velocity fields  are shown in figure \ref{v50_fit}.

\begin{figure}[h!] 
  \centering
   \includegraphics[scale=0.65] {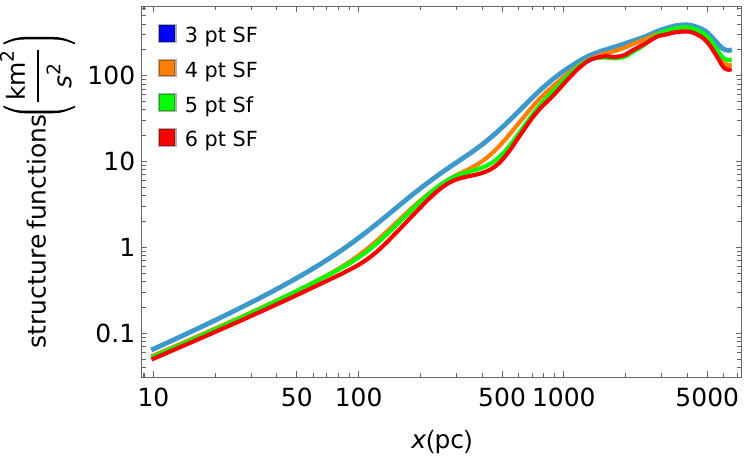 }
  \caption {Multi-point structure functions of the residual  outflowing wind velocity field, in units of (km/s)$^2$ as function of spatial lags along the major axis, in units of    pc. } 
    \label{SFs_v50} 
  \end{figure}

The computed multi point structure functions of this velocity field are displayed in  figure \ref{SFs_v50}.

  Inspection of  figure \ref{SFs_v50} reveals that the 3 point structure function shows only a hint for the small scale turbulence. The 4, 5 and 6 point structure functions all show a common break followed by a plateau. 
 
The  break in  figure \ref{SFs_v50} is  at $l_s\sim$ 290 pc and the plateau has a width of  $\sim$ 130 pc. The  corresponding value of the structure functions is
 \begin{equation}
 SF(l_s)= 2\sigma_s^2= 6.8\ km^2/s^2, 
 \end{equation}
  implying       $ \sigma_s$=1.84 km/s .
  
 The associated  timescale is $l_s/ \sigma_s\sim$ 155  Myr.

   There is an additional break at $l_s \sim$1.5 kpc with a plateau width of $\sim$500 pc. It is evident only in the 5-point and 6-point structure functions.
   
    The corresponding velocity is
 $\sim$9.1\ km/s and the associated timescale is  $\sim$162\ Myr.  This timescale 
   is similar to that corresponding to the smaller break at $l_s\sim$ 290 pc.

 \section{Discussion}
 
We utilized multi-point structure functions to study small scale turbulence in this galaxy, As a first step we obtained interpolation  fits to each of the discrete velocity fields: the residual nebular gas velocity and the residual outflowing wind velocity. Then, the multi-point structure functions for each were obtained, using integrations of the interpolated   velocity fields to evaluate the averages appearing in the definitions of the multi point structure functions. 

The observational data of \citet{Vasan+2025} consist  of line of sight velocities as function of position along the galaxy major axis. We assume that the 3D turbulence is isotropic.This justifies using the structure function formalism for 1D data along a line.
\vskip 0.1 cm

We find  that {\it in addition to} the large scale turbulence studied by \citet{Goldman2024} there exists also a small scale turbulence. The multi-point structure functions of the nebular velocity,  displayed in figure \ref{SFs_v_neb}indicate the existence a small scale turbulence with largest scale $l_s\sim$240 pc, followed by a plateau with a width of $\sim$120 pc.  The small scale  turbulent velocity on this scale i$ \sim$1.86 km/s and the associated timescale is $\sim$ 127 Myr.

The large scale  turbulence velocity at its the largest scale ($\sim$6.43 kpc)   is   (\citet{Goldman2024})  $\sim$13.9 km/s. For wavenumbers between $k_0$ and 1.5 $k_0$ the power spectrum is $\propto k^{-2}$, where $k_0$ is the smallest wavenumber corresponding to the largest spatial scale.  For larger wavenumbers the power spectrum is  $\propto k^{-3}$.  
 
From the above we estimate the turbulent velocity of the large scale turbulence at $l_s$ to be $\sim$0,67 km/s.

 The small scale turbulence velocity at the spatial scale $l_s$ is thus a factor of $\sim$  2.9 larger than that of the large  scale turbulence at $l_s$, and the associated timescale is 2.9 shorter than that of the large scale turbulence at $l_s$.
 
 We turn now to the the case of the wind residual velocity. The multi-point structure functions are displayed in figure \ref{SFs_v50}. One sees that  $l_s\sim$ 290 pc with a plateau of width $\sim$130 pc. The turbulence small scale turbulence velocity on this scale is $\sim$1.84 km/s. The corresponding timescale is thus $\sim$155 Myr. 
 
 The turbulent velocity of the large scale turbulence at this $ l_s$  is evaluated  in a similar way as above. The value of the turbulent velocity on the largest scale  is  (\citet{Goldman2024})  12,1  km/s. Thus, the large scale turbulence velocity at $l_s\sim$290  pc is $\sim$0.67 km/s, a factor of $\sim$2.75 smaller than the small scale turbulence at this $l_s$.
 
The 5-point and 6-point structure functions of figure \ref{SFs_v50} exhibit  an additional break at 
$l_s \sim$1.5 kpc with a plateau width of $\sim$500 pc. The small scale turbulence velocity at this break is
 $\sim$9.1\ km/s and the associated timescale is  $\sim$162 Myr; a value close to that of the smaller  $l_s\sim$290 pc. 
 
 The estimated value of the large scale turbulence velocity at this scale is
$\sim$3.5 km/s; smaller by a factor of $\sim$2.6 than the small scale turbulence velocity.

The decrease in the multi-point structure functions in figures  \ref{SFs_v_neb} and \ref{SFs_v50} at lags $\sim$(5-6) kpc are due to the proximity to  the edge of the $x$ axis and are not break points.
 \vskip 0.5 cm
 The data of \citet{Vasan+2025} dose not include the uncertainties of the two velocity fields. If they would have been available, the uncertainties of the structure functions could have been evaluated in the following procedure. Generating, for example,  1000 simulated     velocity fields in which the velocity at a given position is    randomly displaced from the observational velocity  and is within the observational uncertainty. The structure function  of each of the simulated  velocity fields would be  computed and the   standard deviation of the resulting ensemble of   structure functions would be  taken as the  uncertainty of the structure function. 
 \vskip 0.5 cm

The large spatial scale of the large scale turbulence studied by \citet{Goldman2024} as well  as its large timescale suggest that it has been generated by a merger or tidal interaction that were followed by intense star formation that could have been assisted  also by the large scale compressible turbulence itself  \citep{Gerrard+federrath2026}.

The small scale turbulence, is probably part of the feedback to  the star formation. The values of $l_s\sim$240 pc and $l_s\sim$290 pc suggest  large star clusters or sub-clumps of the large clumps observed in this galaxy, as  generators of the small scale turbulence, probably by intens stellar winds  from massive stars.

The small scale turbulence with  $l_s \sim$1.5 kpc has a largest scale comparable to the size of the large clumps in that galaxy. This  suggests that the latter are the drivers of this turbulence,

The timescale of this small scale turbulence is very close to that with $l_s \sim$290 pc, implying that the two were generated at roughly the same time.

An interesting question is wether the large and small scale turbulence interact. The fact that the timescales of the small scale turbulence are smaller by a factor of ~2 than those of the large scale turbulence at the break points, suggests that there is no interaction.  

The timescale of the large scale turbulence on the largest spatial scale is $\sim$500 Myr; implying that this is also the look-back time to its generation event. The timescale of the small scale turbulence is $\sim$150 Myr, The timescale associated with the large scale velocity gradients are $\leq\sim$50 Myr. This is also the age of the young stars. 

The emerging picture is that the small scale turbulence as well as the velocity gradients were generated on the background of the large scale turbulence. 
          
 \section*{acknowledgment}
 
Thanks are due to the Research authority of Afeka College for support  and to the referee for helpful comments.

   \end{document}